\documentclass{emulateapj}

\newcommand{\citepeg}[1]{\citep[{e.g.,}][]{#1}}
\def\Swift{{\textit{Swift}}\,}

\newcommand{\OII}{[O~II]} 
\newcommand{\OIII}{[O~III]} 
\newcommand{\NII}{[N~II]} 
\newcommand{\SII}{[S~II]} 

\newcommand{\pasa}{PASA}
\newcommand{\actaa}{Acta Astronomica}

\shorttitle{The LIRG Host of Short GRB 100206A}
\shortauthors{Perley et al.}

\begin{document}

\title{The Luminous Infrared Host Galaxy of Short-Duration GRB 100206A}

\def\cit{1}
\def\ucb{2}
\def\nyu{3}
\def\asu{4}
\def\hubble{5}
\def\mail{*}

\author{D.~A.~Perley\altaffilmark{\cit,\ucb,\hubble,\mail},
 M.~Modjaz\altaffilmark{\nyu},
 A.~N.~Morgan\altaffilmark{\ucb},
 S.~B.~Cenko\altaffilmark{\ucb}, \\
 J.~S.~Bloom\altaffilmark{\ucb},
 N.~R.~Butler\altaffilmark{\asu},
 A.~V.~Filippenko\altaffilmark{\ucb}, and 
 A.~A.~Miller\altaffilmark{\ucb}
}

\altaffiltext{\cit}{Department of Astronomy, California Institute of Technology, MC 249-17, 1200 East California Blvd., Pasadena, CA 91125, USA.}
\altaffiltext{\ucb}{Department of Astronomy, University of California,  Berkeley, CA 94720-3411, USA.}
\altaffiltext{\nyu}{Center for Cosmology and Particle Physics, New York University, 4 Washington Place, New York, NY 10003, USA}
\altaffiltext{\asu}{School of Earth and Space Exploration,  Arizona State University, Tempe, AZ 85287, USA.}
\altaffiltext{\hubble}{Hubble Fellow.}
\altaffiltext{\mail}{e-mail: dperley@astro.caltech.edu .}

\slugcomment{Submitted to ApJ 2011-12-16}

\begin{abstract}
The known host galaxies of short-hard gamma-ray bursts (GRBs) to date are characterized by low to moderate star-formation rates and a broad range of stellar masses.  In this paper, we positionally associate the recent unambiguously short-hard \Swift\ GRB 100206A with a disk galaxy at redshift $z=0.4068$ that is rapidly forming stars at a rate of $\sim 30$ M$_{\odot}$ yr$^{-1}$, almost an order of magnitude higher than any previously identified short GRB host.   Using photometry from Gemini, Keck, PAIRITEL, and WISE, we show that the galaxy is very red ($g-K = 4.3$ AB mag), heavily obscured ($A_V \approx 2$~mag), and has the highest metallicity of any GRB host to date (12 + log[O/H]$_{\rm KD02}$ = 9.2): it is a classical luminous infrared galaxy (LIRG), with $L_{\rm IR} \approx 4 \times 10^{11} {\rm L}_{\odot}$.  While these properties could be interpreted to support an association of this GRB with very recent star formation, modeling of the broadband spectral energy distribution also indicates that a substantial stellar mass of mostly older stars is present.  The current specific star-formation rate is modest (specific SFR $\approx$ 0.5 Gyr$^{-1}$), the current star-formation rate is not substantially elevated above its long-term average, and the host morphology shows no sign of recent merger activity.   Our observations are therefore equally consistent with an older progenitor, similar to what is inferred for other short-hard GRBs.  Given the precedent established by previous short GRB hosts and the significant fraction of the Universe's stellar mass in LIRG-like systems at $z \gtrsim 0.3$, an older progenitor represents the most likely origin of this event.
\end{abstract}


\section{Introduction}
\label{sec:intro}

Classical gamma-ray bursts (GRBs) segregate into two phenomenological classes on the basis of their prompt-emission properties: long-duration, soft-spectrum GRBs and short-duration, hard-spectrum GRBs \citep{Kouveliotou+1993}.  The two distributions overlap in both duration and in hardness, but a duration of $T_{90} = 2$~s is commonly taken as the dividing line between ``long'' and ``short'' bursts\footnote{However, the question of classification has recently become much more complex following the discovery of events with extended emission (\citealt{Norris+2006}) and deep limits on supernovae from $t>2$~s bursts (\citealt{Fynbo+2006,Gehrels+2006,GalYam+2006}).  See \cite{Bloom+2008} for a discourse on the subject of classification.}.  Long bursts constitute the large majority of events detected by all major GRB satellites (about 75\% of BATSE events and over 95\% of \Swift\ events); because these events have brighter and longer-lived afterglows \citep{Kann+2011}, they have historically been easier to study.   Observations stretching back almost 15 years associate long-duration bursts exclusively with actively star-forming host galaxies, and specifically with regions bright at ultraviolet (UV) wavelengths \citep{Fruchter+2006} close to the center \citep{Bloom+2002} of these galaxies, strongly pointing toward a massive stellar origin.  In a number of cases, evidence for an accompanying supernova has provided direct evidence confirming this conclusion (see \citealt{WoosleyBloom2006} for a review).

The short GRB sample is much smaller, but its properties are unambiguously different from those of long bursts.  At least two short events (GRBs 050509B and 050724; \citealt{Bloom+2006,Gehrels+2005,Barthelmy+2005,Berger+2005,Prochaska+2006}) have been associated with massive, evolved galaxies having essentially negligible current star formation, pointing strongly toward a long-lived progenitor \citep{BloomProchaska2006}.   In addition, the offset distribution for short GRBs relative to their host galaxies extends to much greater distances than for long GRBs --- partially as a simple result of the fact that short-GRB hosts tend to be physically more extended \citep{Fong+2010}, but short GRBs also generally show minimal correlation with the host blue/UV light or resolved regions of active star formation \citep{Fox+2005,Fong+2010,Rowlinson+2010b}.   The occurrence of an accompanying bright core-collapse supernova has been definitively ruled out for several short GRBs \citepeg{Hjorth+2005a,Kocevski+2010}. 

The model of short GRBs as mergers of compact binary stars \citep{Eichler+1989,Paczynski+1991,Narayan+1992} --- either two neutron stars (NS-NS) or a neutron star and a black hole (NS-BH) --- naturally predicts old progenitors, 
and the observations discussed above appear to strongly support this model.   However, other models permitting older populations do exist \citepeg{MacFadyen+2005}, and recent observations have painted a picture that appears more complex than it initially seemed after the discovery of the first few short-GRB hosts.  A significant fraction of short GRBs with afterglows actually have no clearly identifiable host \citep{Berger2010}, suggesting that they have been flung large distances from their original hosts or occur in optically underluminous, distant galaxies in surprisingly large numbers.  Finally, with the exceptions of GRBs 050509B, 050724, and possibly 100117A \citep{Fong+2011}, every short GRB host to date (at this stage, a fairly formidable sample of $>$20 objects) has had at least modest star formation in relation to its stellar mass \citepeg{Berger2010}, making it difficult to pin down a minimum progenitor age for the large majority of these events.  The two ``smoking gun'' events of 2005 no longer appear to be particularly representative of the entire class.

The relationship between short and long GRBs is at least vaguely analogous to that between Type Ia and Type II supernovae \citep[e.g.,][]{Wheeler1981}: Type Ia supernovae are often associated with moderate or old populations (and almost certainly are produced by compact objects, in this case white dwarfs; \citealt{Nomoto1982,Bloom+2011b,Nugent+2011}), while Type II supernovae are associated exclusively with young populations (and result from massive stellar core collapse).  
However, recent evidence suggests that Type Ia supernovae can actually be produced fairly rapidly in some cases, given their elevated rate in late-type hosts compared to what might be expected from an exclusively old population \citep{ScanBildsten2005,Mannucci+2005,Sullivan+2006,Li+2011,Maoz+2011} --- that is, the delay-time distribution is likely to be quite broad, with short-lived stellar progenitors as well as long-lived ones.

It is possible that the short GRB delay-time distribution may be similarly complex.  Confirming this trend --- and in particular, determining whether there is any need for a very young component associated with the youngest (and presumably most massive) stars, which could indicate a physically distinct progenitor \citepeg{Metzger+2008,Lazzati+2010,Virgili+2011} --- is of clear interest for better understanding the origins of the short-GRB population.

Given the low short-GRB event rate ($< 10$ events yr$^{-1}$) and uncertain selection biases affecting the population ($\sim 25$\% of all \textit{Swift} short GRBs are not detected by the XRT and cannot be localized), teasing such a tendency out of the full population is possible but challenging.  Nevertheless, two recent studies have attempted to accomplish this using different techniques.   \cite{Virgili+2011} analyzed the short-GRB redshift, fluence, and luminosity distributions compared to predictions from a delay-time convolved star-formation history of the Universe and concluded that a significant fraction of short GRBs may in fact originate from young stars.  \cite{Leibler+2010} used the properties of the host galaxies of a sample of short GRBs to estimate the age of the predominant stellar population, and similarly concluded that both short-delay and long-delay components are necessitated by the data.  However, even the ``short'' component is consistent with a timescale of order 200 Myr --- much older than the lifetimes of massive stars that produce core-collapse supernovae, long GRBs, and nebular signatures in star-forming galaxies.

The most unambiguous indicator of a very short-delay ($\lesssim$100 Myr) component would be the discovery of another ``smoking gun'' system, but this time with an extremely \emph{young} stellar age.  In particular, a short GRB within a starbursting galaxy whose current star-formation rate (SFR) is very large in comparison to its stellar mass (high specific SFR) would be a strong indicator that the event came from a star produced in the starburst, events whose characteristic times rarely exceed a few hundred Myr and are often much less \citep{DiMatteo+2008,McQuinn+2009,McQuinn+2010}.

In this paper, we discuss the case of GRB 100206A as the first example of a short-duration burst whose host characteristics, at least on the surface, evoke such a system.   In \S \ref{sec:obs} we present our observations of the burst and its afterglow, showing it to be an unambiguous short-duration, hard-spectrum GRB with a faint, rapidly-fading afterglow detected only in the X-ray band despite deep, early optical imaging.  We also present observations of two galaxies in or near the X-ray error circle at a variety of wavelengths from optical through the mid-infrared.  In \S \ref{sec:anal} we analyze these data in further detail, measuring the metallicity and constraining the underlying stellar population of the brighter host candidate, which we show to be a luminous infrared galaxy (LIRG) at redshift $z=0.4068$ with a current SFR exceeding that of any previous short GRB by almost an order of magnitude.  In \S \ref{sec:disc} we argue that {\it a posteriori} statistical arguments strongly tie this galaxy to the short GRB, and examine the implications for the progenitor of the short-duration burst.  While the current SFR is high, we note that the stellar mass is substantial and dominated by older stars, with no unambiguous evidence for starbursting, merging, or other short-lived features --- consistent with an older progenitor, like that of other short-hard bursts.  We summarize our conclusions in \S \ref{sec:conc}.

\section{Observations}
\label{sec:obs}

\subsection{Swift-BAT and Fermi}

\begin{figure}
\centerline{
\includegraphics[scale=0.7,angle=0]{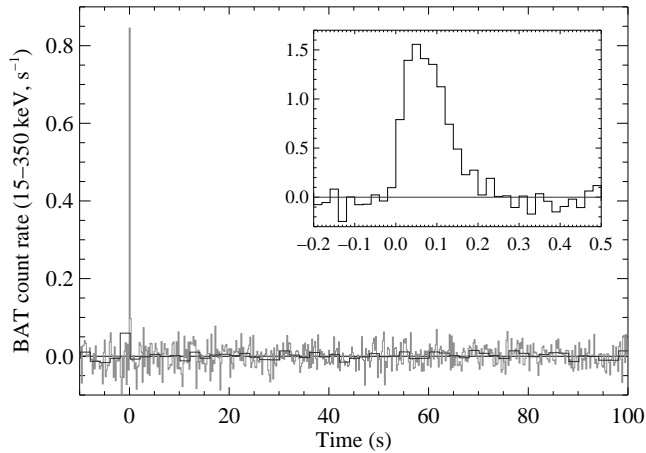}} 
\caption{BAT light curve of GRB 100206A, taken from the sum of all four channels of data (15--350 keV) using the methods of of \citealt{Butler+2007}.  The main plot shows the light curve binned at 2~s (dark gray) and 0.2~s (light gray); the inset shows the light curve binned at 0.02~s.  Times are referenced to $t=$949498223.86 sec (GPS).  The burst is clearly short, with all detectable gamma-ray emission contained within a 0.25~s interval.  There is no evidence of extended emission during the minutes after the trigger.}
\label{fig:batlc}
\end{figure}

GRB~100206A triggered the Burst Alert Telescope (BAT; \citealt{Barthelmy+2005}) on the {\it Swift} satellite \citep{Gehrels+2004} at 13:30:05 on 2011 February 06 (UT dates are used throughout this paper).  BAT data are automatically reduced by our automated pipeline using the methods of \cite{Butler+2007}.  The BAT light curve (Figure \ref{fig:batlc}) shows only a single spike starting at the trigger time and ending by 0.2~s, with no evidence of extended emission (with $T_{90}=0.200\pm0.017$~s; \citealt{GCN10379}). 

The GRB also triggered the Gamma-ray Burst Monitor (GBM; \citealt{Meegan+2009}) onboard the {\it Fermi} satellite.  The {\it Fermi} light curve is similar to that seen by \Swift---a single bright spike lasting $<0.2$~s ($T_{90} = 0.13 \pm 0.05$~s; \citealt{GCN10381}).   The spectrum of the burst is quite hard; the best-fit \cite{Band+1993} model indicates $E_{\rm peak} =  439^{+73}_{-60}$ keV.

These properties place GRB 100206A unambiguously within the short-duration phenomenological class (see, for instance, Figure 1 of \citealt{Levesque+2010_090426}).

\subsection{Swift-XRT and UVOT}

{\it Swift} slewed immediately to the source and began pointed observations with the X-Ray Telescope \citep[XRT;][]{Burrows+2005} at 74.6~s after the BAT trigger, followed by observations with the Ultraviolet Optical Telescope (UVOT; \citealt{Roming+2005}) beginning at 78~s.   Only a faint and rapidly fading X-ray afterglow is seen, dropping from 0.2 to 0.01 count s$^{-1}$ during the first orbit; in total, only 24 X-ray photons were detected by the instrument during this observation.   \Swift\ made several return visits, none of which resulted in a detection above the background level.

During the first orbit, the event data are consistent with a light curve following a sharply falling, unbroken power law with decay index $\alpha = 1.91^{+0.40}_{-0.47}$. Given the limited number of counts and large Galactic hydrogen column ($N_{\rm H} \approx 10^{21}$ cm$^{-2}$), no clear statements can be made about the intrinsic X-ray spectrum.   The best available localization is the UVOT-enhanced XRT position \citep{Goad+2007}: $\alpha = 03^{\rm h}08^{\rm m}39^{\rm s}.03$,$\delta =$ +13\arcdeg 09\arcmin 25.3\arcsec\ (J2000, 3.3\arcsec\ uncertainty at 90\% confidence).

No detection is reported in UVOT observations at any epoch.  UVOT upper limits are given by \cite{GCNR271.1}.  In addition, we stacked all available UVOT $u$-band imaging of this object to try to provide a deep limit on any host-galaxy emission; the integrated limit is $u > 21.8$ mag.  Unfortunately, given the relatively large Galactic extinction in the field ($E_{B-V} = 0.382$ mag; \citealt{Schlegel+1998}), this value is not particularly constraining.

\subsection{Ground-Based Follow-Up Observations}

GRB 100206A was observed by many different ground-based instruments within the first 24~hr after the trigger \citep{GCN10380,GCN10384,GCN10385,GCN10386,GCN10387,GCN10388,GCN10390,GCN10391,GCN10392,GCN10395,GCN10396,GCN10455,GCN10456}.  None of these observations report a detection of a varying source, although two static sources in the XRT error circle, initially noted by \citet{GCN10377} and \citet{GCN10386}, will be discussed below.  Given the faint X-ray afterglow, this nondetection of variability is unsurprising, as explained in further detail in \S \ref{sec:lc}.

\subsection{DeepSky}

\begin{figure}
\centerline{
\includegraphics[scale=0.46,angle=0]{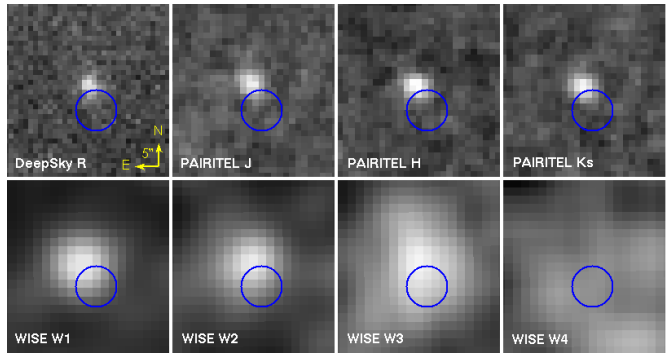}} 
\caption{Unresolved images of the putative host galaxy (G1) of GRB 100206A in the optical and near-infrared (NIR) from the 1.2~m Palomar Oschin Telescope (from the Palomar/DeepSky project), the 1.3~m PAIRITEL, and the WISE all-sky mission.  Although these telescopes are relatively insensitive to typical galaxies at cosmological distances, a bright source centered just outside the XRT error circle is well detected in every filter except W4 (22~$\mu$m).}
\label{fig:bwtile}
\end{figure}

We co-added 78 archival images from the DeepSky project at Palomar Observatory \citep{Nugent+2009} covering the field of GRB 100206A.  The images were obtained in 2004--2008 from the Palomar-Quest Consortium at the Oschin Schmidt telescope.  The limiting magnitude of the stack is $R \approx 23$ mag.  In the combined image we detect a faint ($R = 21.7 \pm 0.3$ mag relative to nearby USNO catalog stars) extended source, centered slightly outside the current XRT error circle to its northeast (Figure \ref{fig:bwtile}).

The random appearance of a source this bright close to a small X-ray error circle
is small, but not necessarily convincingly so  ($P_{\rm chance} \approx 0.05$; see \S \ref{sec:pchance}).
However, this source has very unusual properties at other wavelengths that show it is far from being a typical galaxy.

\subsection{PAIRITEL}
The robotic Peters Automatic Infrared Imaging Telescope (PAIRITEL; \citealt{Bloom+2006}) began observations of the GRB 100206A field starting 13.1~hr after the burst in the $J$, $H$, and $K_s$ filters simultaneously.  No transient source was detected inside the error circle.

However, the host-galaxy candidate first seen in the DeepSky  archival images was well detected in all three filters.  PAIRITEL detection of galaxies at cosmological distances (\S \ref{sec:keckspec}) is unusual, leading us to first speculate that this galaxy may be a LIRG \citep{GCN10390}.  To improve the photometry of this source and verify the lack of fading behavior, we acquired additional imaging on 2011 October 20 and 2011 October 22, and combined data from all three epochs (total effective exposure time of 4.34~hr) with the exception of about 2~hr of $K_s$-band data which was not usable due to poor sky conditions.  

Aperture photometry was performed using custom Python software, utilizing Source Extractor (SExtractor; \citealt{Bertin+1996}) as a back end.  Calibration was performed by determining the zeropoint for each image by comparison to 2MASS \citep{Skrutskie+2006} magnitudes using 19 calibration stars.  As the galaxy is clearly extended in higher-resolution images, we employ a large aperture radius of 3\arcsec.  

Indeed, no evidence of fading (to limits of $J > 19.8$, $H > 19.3$, $K_s > 16.8$ mag) is observed between the epochs (the $K_s$-band constraint is poor due to the low quality of the second-epoch image).  Photometry of the host galaxy using the final, combined stacks is presented in Table \ref{tab:hostphot}.

\begin{deluxetable*}{l|ll|ll|ll}
\tabletypesize{\small}
\tablecaption{Photometry of Galaxies Inside the XRT Error Circle of GRB\,100206A\label{tab:phot}}
\tablecolumns{7}
\tablehead{
\colhead{Instrument} &
\colhead{Exp.~time} &
\colhead{Filter} &
\colhead{Magnitude\tablenotemark{a}} & 
\colhead{AB mag.\tablenotemark{b}} &
\colhead{Magnitude\tablenotemark{a}} & 
\colhead{AB mag.\tablenotemark{b}} 
\\
\colhead{} &
\colhead{(s)} & 
\colhead{} & 
\multicolumn{2}{c}{G1 ($z=0.41$)} & 
\multicolumn{2}{c}{G2 ($z=0.80$)}
}
\startdata
Keck I / LRIS     &  600  & $g$  & 23.83$\pm$0.17 & 22.45  &  26.55$\pm$0.40 & 25.17   \\
Keck I / LRIS     &  540  & $R$  & 21.31$\pm$0.09 & 20.51  &  25.14$\pm$0.19 & 24.34   \\
Gemini-N / GMOS-N & 2400  & $i$  & 20.82$\pm$0.08 & 20.06  &  24.69$\pm$0.05 & 23.93   \\
Gemini-N / GMOS-N & 1200  & $z$  & 20.20$\pm$0.05 & 19.63  &  24.15$\pm$0.12 & 23.57   \\
PAIRITEL          & 16333 & $J$  & 18.51$\pm$0.12 & 19.06  &  $>$19.7        & 20.3    \\
PAIRITEL          & 16333 & $H$  & 17.25$\pm$0.09 & 18.41  &  $>$18.8        & 19.9    \\
PAIRITEL          & 8143  & $K_s$& 16.33$\pm$0.11 & 18.03  &  $>$17.7        & 19.4    \\
WISE              &       & W1   & 15.74$\pm$0.06 & 18.42  &                 &         \\
WISE              &       & W2   & 15.14$\pm$0.11 & 18.47  &                 &         \\
WISE              &       & W3   & 11.23$\pm$0.16 & 16.38  &                 &         \\
WISE              &       & W4   &$>$8.58         &$>$15.18&                 &         \\
\enddata 
\tablenotetext{a}{Observed value, not corrected for Galactic extinction.}
\tablenotetext{b}{Corrected for Galactic extinction ($E_{B-V} = 0.38$ mag).}
\label{tab:hostphot}
\end{deluxetable*}

\subsection{Gemini Imaging}

Two epochs of deep imaging were acquired with the Gemini Multi-Object Spectrometer (GMOS-N) on Gemini-North.  In the first epoch (starting at 05:34, 14~hr after the trigger), five dithered exposures of 240~s each were obtained in the $z$ band, immediately followed by a similar sequence of $5 \times 240$~s exposures in the $i$ band.    The $i$-band sequence was repeated five days later, between 06:41 and 07:05 on 2010 February 12.  Conditions were excellent during both observations, with an average seeing of about 0.8\arcsec.

\begin{figure}
\centerline{
\includegraphics[scale=0.4,angle=0]{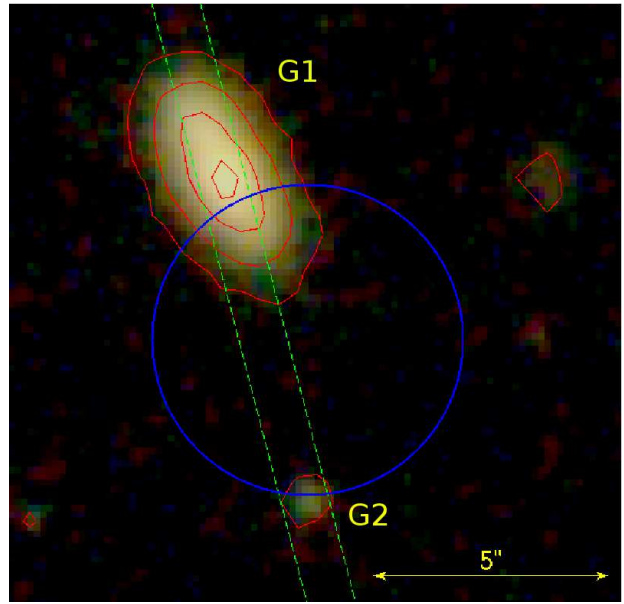}}  
\caption{Deep imaging of the field of GRB 100206A from the Keck 10~m telescope and Gemini-North 8~m telescope, combined into a false-color image using the $R$, $i$, and $z$ filters.  The bright galaxy G1 resolves into a highly inclined disk, with most of the southern half of the galaxy enclosed by the XRT error circle (shown in blue).  A second, much more compact galaxy is also evident, centered at the southern edge of the XRT error circle, which we label G2.  Isophotes of G1 show a subtle asymmetry that probably results from additional substructure within the galaxy (see also Figure \ref{fig:deconv}.)  The dashed green line indicates the position of the LRIS slit during our long-slit spectroscopy.  North is up and East is left.}
\label{fig:colorimg}
\end{figure}

The bright archival galaxy first seen in our DeepSky image is now resolved into an extended disk of about 5\arcsec\ diameter (Figure \ref{fig:colorimg}).  We denote this galaxy ``G1'' in the remainder of the paper.   A second, much fainter point-like source (which we denote ``G2'') is also seen $\sim 8$\arcsec\ to the south of this object.

\cite{GCN10395}, who triggered these observations, performed image subtraction between the two epochs of $i$-band imaging and reported no variation either between these epochs or in comparison with a WHT image taken 7~hr after the trigger --- including, in particular, no variation of the point-like source G2 which had been initially suggested by \cite{GCN10386} as a candidate afterglow.  (Indeed, our spectroscopy verifies that this object is a background galaxy; \S \ref{sec:keckspec}.)

We downloaded all Gemini frames from the Gemini Science archive and reduced them independently using the GMOS IRAF reduction tools.  We first coadded the data obtained during each of the two separate $i$-band epochs, and subtracted the two images using HotPants\footnote{\url{http://www.astro.washington.edu/users/becker/hotpants.html .}}, confirming the lack of variability reported by Berger et al.\footnote{One marginal source does appear with a nominal significance of about 5$\sigma$ at $\alpha = 03^{\rm h}08^{\rm m}38.846^{\rm s}$, $\delta =$ +13\arcdeg 09\arcmin 24.25\arcsec\ (J2000) but it seems to be mostly due to a weak negative artifact in the late-time reference image rather than the detection of positive variation in the first image.}  We measure an improved 5$\sigma$ upper limit of $i > 25.9$ mag for any point source varying between the two frames over an 0.8\arcsec\ aperture within the XRT error circle.

Given the lack of variation and comparable, good conditions over the two nights, we then stacked all ten $i$-band exposures over both epochs to produce a single, deep image.  The five $z$-band frames from the first night were also coadded into a separate $z$-band stack.  Using secondary standard stars from our P60 calibration of the field (\S \ref{sec:calib}), we performed aperture photometry of G1 using a radius of 3\arcsec.  Photometry of G2 was calculated using the same standards, using a 1\arcsec\ radius aperture; see Table \ref{tab:hostphot}.

\subsection{Keck Imaging}
\label{sec:keckim}

To obtain additional color information, on the night of 2010-08-02 we imaged the field of GRB\ 100206A with the Low Resolution Imaging Spectrograph \citep[LRIS;][]{Oke+1995} on the 10~m Keck~I telescope.   Using the D560 dichroic, we acquired images in the $g$ and $R$ filters simultaneously, with $3 \times 200$~s integration in $g$ and $3 \times 180$~s integration in $R$. Reduction was accomplished using a custom LRIS imaging pipeline. As with the Gemini images, we performed aperture photometry of the two possible host galaxies G1 and G2 using 3\arcsec\ and 1\arcsec\ radius apertures, respectively, using our secondary field standards.  Again, the photometry is reported in Table \ref{tab:hostphot}.

\subsection{P60 and Nickel Photometric Calibrations}
\label{sec:calib}

To calibrate the optical photometry, we acquired independent calibrations using the roboticized 60-inch telescope at Palomar Observatory (P60; \citealt{Cenko+2006}) and the Nickel 1~m reflector at Lick Observatory, both on 2011 August 09.  Numerous standard fields from \cite{Landolt2009} were observed throughout the night in the $gRiz$ filters (P60) or $BVRI$ filters (Nickel).  The Nickel photometry was then transformed to $giz$ using the transformation equations of \cite{Jester+2005}.  The two calibrations show good consistency, with $< 0.04$ mag systematic differences in each of the $Riz$ filters for bright stars.  A slightly larger offset of $0.1$ mag is observed in the $g$ band.  To calibrate the galaxy photometry, we take the average of the two $g$, $R$, and $i$-band calibrations, adding a small calibration component to the uncertainty.  Only the P60 data are used to calibrate the $z$ band.  Final, calibrated photometry of the two galaxies near the XRT error circle is presented in Table \ref{tab:phot}.

\subsection{Keck Spectroscopy}
\label{sec:keckspec}

We acquired spectra using LRIS on the Keck~I 10~m telescope on two occasions, both using the D560 dichroic, the 600/4000 grism, the 400/8500 grating, and a 1\arcsec\ slit. The first epoch was obtained on 2010 February 07, shortly after the burst.  Two exposures, each of duration 600~s (on the red side; slightly longer on the blue side), were acquired at a slit position angle (PA) of $18.2^\circ$ between 06:17 and 06:39.  (LRIS is equipped with an Atmospheric Dispersion Corrector [\citealt{Phillips+2006}], enabling us to observe away from the parallactic angle without slit losses.)  Seeing conditions were poor ($\sim 1.5$\arcsec) throughout the integration.   Nevertheless, the galaxy was well detected in the red part of the spectrum, showing strong emission lines of H$\alpha$ and \NII, as we previously noted \citep{GCN10389}.  The slit orientation was fortuitously close to the galaxy major axis and these lines show clear rotational structure.

A second epoch was acquired on 2011 August 02, shortly following our imaging that same night (\S \ref{sec:keckim}).   A slit PA of $14.5^\circ$ was used to simultaneously cover both G1 and G2 at the same time; one exposure was acquired of 900~s duration (on the red side; the integration on the blue side lasted an additional 10~s).  This PA is still close to the major axis of the galaxy (Figure \ref{fig:colorimg}).   Conditions were excellent (0.8\arcsec\ seeing) and the spectrum is of significantly higher quality than the one obtained during the earlier epoch, showing both lines of the \NII\ doublet as well as \SII, H$\beta$, and \OII.  Accordingly, only this spectrum is used in our final analysis.

The spectra were reduced using custom reduction software written in IDL.  The traces of the two galaxies were extracted separately using our custom software and flux-calibrated relative to spectroscopic standards BD+284211 (blue; \citealt{Oke1990}) and BD+174708 (red; \citealt{OkeGunn1983}), and by comparison of synthetic photometry of the spectrum to the photometry from our images.  Both objects are observed to have multiple strong emission lines, identifying them as galaxies at redshifts of $z=0.4068$ (G1) and $z=0.803$ (G2).   In the case of G1, strong rotational ``shearing'' is seen in the two-dimensional spectrum ($v/c = 0.0014$ between opposite ends of the galaxy, corresponding to 13~\AA\ in the vicinity of H$\alpha$), so a simple extraction would produce much poorer resolution than what is provided by the instrument.  To remove this shear, we calculate the rotation curve of the galaxy by finding the maximum of the cross-correlation function along each row of the trace relative to the galaxy center in the region of the strong lines of the H$\alpha$--\NII\ complex, and then resample each line (across the full trace) by the appropriate amount to remove the systematic rotation.  This procedure allows us to achieve a resolution comparable to the instrumental resolution of $\sim 7$~\AA.   Following this step, the spectrum is extracted normally using a window of 3.24\arcsec\ (diameter).  (The emission lines of G2 are not resolved, and this spectrum is extracted normally with no additional steps.)  The final 1D spectra of G1 and G2 are presented in Figures \ref{fig:spectrumg1} and \ref{fig:spectrumg2}, respectively.

\begin{figure}
\centerline{
\includegraphics[scale=0.55,angle=0]{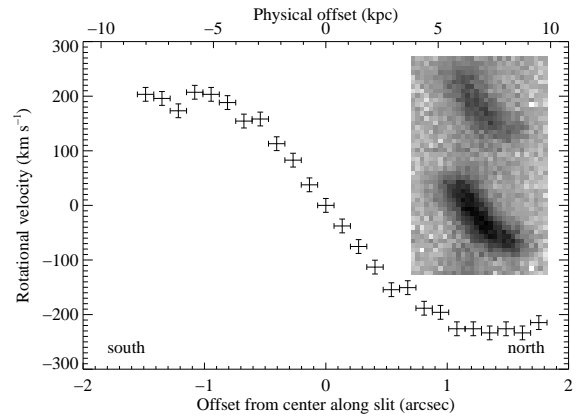}}  
\caption{Rotation curve for G1, determined from H$\alpha$ and \NII\ emission lines.  Note that the slit is 1\arcsec\ (5.4 kpc) wide and at a (slightly) different orientation than the major axis of the galaxy, so the true major-axis rotation velocities of the galaxy are likely to be somewhat larger than the values indicated here.   (The galaxy itself is also likely inclined $\sim 20^\circ$ from edge-on; Figure \ref{fig:deconv}.) The relevant region of the two-dimensional spectrum is shown in the inset.}
\label{fig:rotcurve}
\end{figure}

\begin{figure*}
\centerline{
\includegraphics[scale=0.76,angle=0]{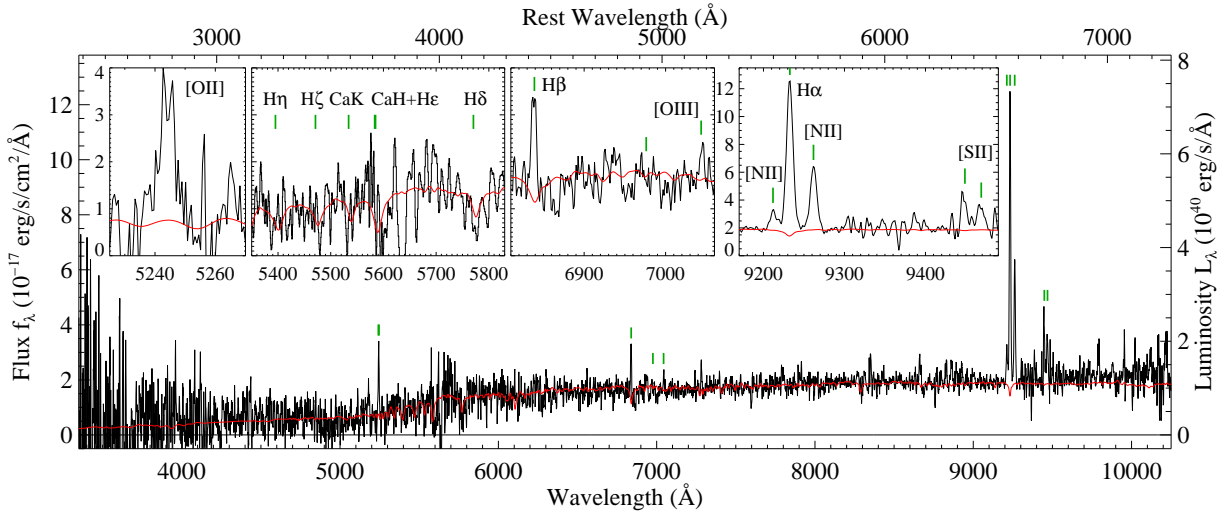}}  
\caption{Spectroscopy of G1, the putative host galaxy.  The combined LRIS blue+red spectrum is shown in the main panel, with the positions of detected emission lines indicated.  In addition, the red curve shows the best stellar-continuum model from a fit to our broadband (and synthetic narrowband) photometry.  The small deviation between this model and the observations at $>$9500~\AA\ is probably due to uncertainties in the spectrophotometry in the long-wavelength region.  At top, the insets show regions around specific lines, including \OII, the Balmer absorption lines, H$\beta$+\OIII, and H$\alpha$+\NII+\SII.}
\label{fig:spectrumg1}
\end{figure*}

\begin{figure*}
\centerline{
\includegraphics[scale=0.76,angle=0]{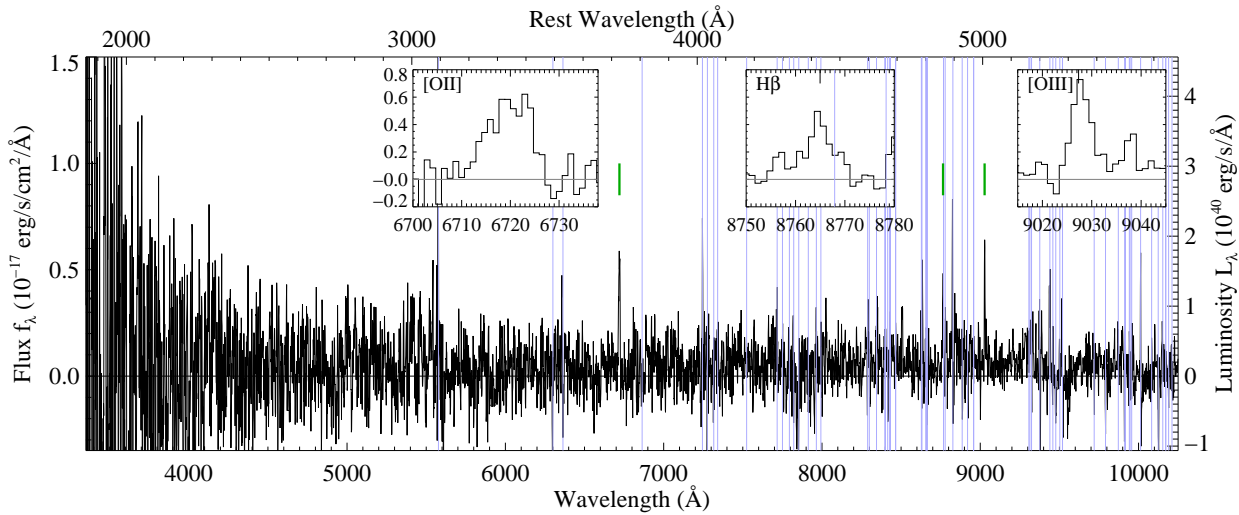}}  
\caption{Spectrum of G2, the faint galaxy at the south end of of the XRT error circle.  Secure detection of the \OII\ and \OIII\ lines show this to be a star-forming galaxy at $z=0.803$.  The vertical gray lines indicate the centers of strong sky emission lines.}
\label{fig:spectrumg2}
\end{figure*}

\subsection{WISE Archival Observations}
\label{sec:wise}

We searched the Wide-Field Infrared Survey Explorer (WISE; \citealt{Wright+2010}) archive for observations at the position of this GRB.  The field was covered by the WISE preliminary data release, and a source is clearly detected at the position of G1 in both the archival images and in the photometric catalog in filters W1--W3 (no detection is evident in W4), as shown in Figure \ref{fig:bwtile}. In Table \ref{tab:hostphot} we present catalog magnitudes of this object from the archive.

\section{Analysis}
\label{sec:anal}

\subsection{Afterglow and Kilonova Models}
\label{sec:lc}

Despite extensive optical and NIR follow-up observations during the first night (including some data on time scales as short as minutes and a very deep image with Gemini), no afterglow was detected from this event.  Using standard assumptions about the intrinsic spectral index of a GRB afterglow, we examined whether these limits might usefully constrain the properties of the GRB or the extinction column.

\begin{figure}
\centerline{
\includegraphics[scale=0.7,angle=0]{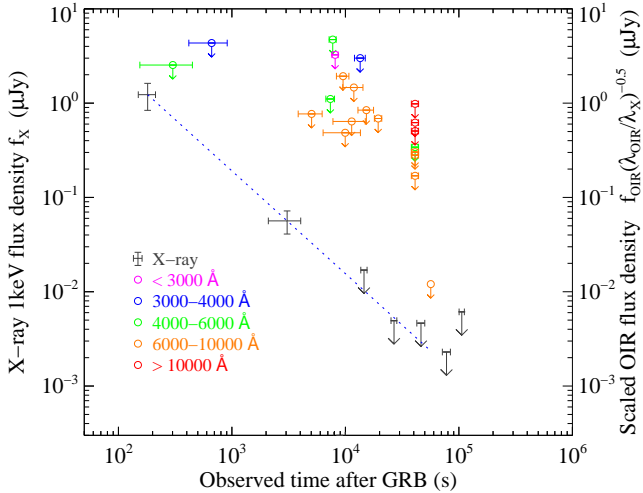}} 
\caption{X-ray and UV/optical/NIR observations of GRB 100206A.  The X-ray light curve is plotted as flux density ($f_{\nu}$) at 1~keV.  To meaningfully place the UV/optical/NIR observations (all of which are upper limits) on the same plot, we calculate the flux corrected for Galactic extinction and extrapolate it into the X-ray band assuming $\beta = 0.5$ ($f_\nu = \nu^{-\beta}$).  Given the extremely faint X-ray afterglow, none of the optical limits is constraining.}
\label{fig:xrtoptlc}
\end{figure}

In Figure \ref{fig:xrtoptlc} we plot the ``light curve'' of the X-ray afterglow (consisting of two binned detections followed by upper limits) and scale all reported UV, optical, or NIR upper limits from the GCN Circulars to the X-ray band (1~keV equivalent photon energy) based on an assumed afterglow spectral index of $\beta_{\rm OX}=0.5$, the minimum value expected in the synchrotron afterglow model (\citealt{Jakobsson+2004a}; a value below this line would indicate a ``dark'' burst).  Any upper limits lying below the X-ray light curve are inconsistent with this basic model and would require either an unusually blue/hard intrinsic spectral index or extinction within the host galaxy to suppress the optical flux.  As is evident from the figure, none of the limits constrain the afterglow in this way.  In fact, given that the second X-ray point is a nondetection, even if we assume a much softer/redder spectral index (up to $\beta_{\rm OX} = 1.1$, which corresponds to an unbroken spectral index between the X-ray and optical given typical X-ray spectral indices and represents the maximum value expected in the synchrotron model), the optical photometry still imposes no constraint on any additional host extinction.

The time of the deep Gemini limit is better timed to constrain emission from a \cite{LiPaczynski1998} mini-supernova or ``kilonova'' (see also \citealt{Kulkarni2005,Metzger+2010b}), which is expected to peak at $t \approx 1$ day.  After correcting for Galactic extinction, the GMOS $i$-band limit corresponds to a specific luminosity limit of $\nu L_{\nu} < 10^{42}$ erg s$^{-1}$ at $t = 11$~hr, if the GRB is at the redshift of the brighter galaxy ($z=0.4068$) and is unextinguished by dust within its host.  This is comparable to the limits presented by \cite{Kocevski+2010} and broadly rules out models with large energy conversion factors and ejecta masses ($f \gtrsim 10^{-5}$ and $M \gtrsim 10^{-2.5} M_{\odot}$), but is significantly less constraining than the limits presented on radioactive-powered emission from GRB 050509B by \citet{Hjorth+2005a}.  Furthermore, even our weak limit is subject to the strong caveat that the extinction within this host galaxy appears to be quite large (\S \ref{sec:sed}) and the assumption of no extinction may not be valid.

\subsection{Morphology}
\label{sec:morphology}

The galaxy G1 has a disk-like morphology (Figure \ref{fig:colorimg}), and is seen at a high inclination.  It is also quite large, with significant emission extending out 1.5\arcsec\ (8 kpc) in each direction away from the nucleus.  Some subtle asymmetry is visible: the location of maximum flux is displaced by about one pixel (0.15\arcsec) from the center of the outer isophotes, and there is a hint of displacement between the northern and southern sides of the disk from the nucleus, possibly suggestive of spiral or bar structure.

To gain further insight into the structure of the galaxy, we used the software package GALFIT \citep{Peng+2002} to model the system as an inclined \cite{Sersic1963} disk.  Our modeling is performed using the Gemini GMOS $i$-band image, which has the best signal-to-noise ratio and seeing quality of the filters used.  A bright, isolated, nonsaturated, nearby star in the image is used as the point-spread function (PSF) model.  Because a simple Sersic fit gives significant residuals (as expected, given the visual asymmetry), we extend this basic model by successively adding individual Gaussians initially centered at locations of positive residuals and iteratively repeat the fit until no residuals are evident above the background noise.  We found three such components to be needed within the physical extent of the disk: one each on either side of the nucleus in the disk plane, each at a displacement of 1.3\arcsec\ (7 kpc in projection), plus an additional component displaced 1.1\arcsec\ east (6 kpc) of the nucleus.  (A fourth component 3.2\arcsec\ northeast of the galaxy is also required by the fitting procedure, but is probably not related to the system.)  These additional components combined only contribute $\sim 20$\% of the total flux of the system, which is still dominated by the Sersic disk.

Whether or not these additional parameters are added, similar results for the Sersic parameters are achieved (Sersic index $n$ = 0.58, and axis ratio $b/a = 0.31$ for a fit with the disk only and no extra components; $n = 1.27$, $b/a = 0.40$ for the final model with four extra components).  Evidently, the stellar light in this system follows a nearly exponential disk ($n=1$; typical for disk galaxies), inclined at an angle of $\sim 70^\circ$ from face-on.  

\begin{figure}
\centerline{
\includegraphics[scale=0.45,angle=0]{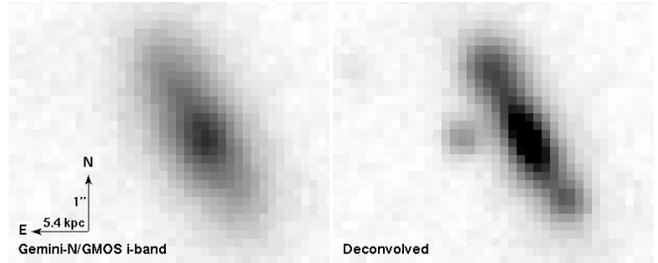}}  
\caption{The field of GRB 100206A imaged in the $i$ band using the Gemini-North 8~m telescope, with no additional processing (left) and as deconvolved by our GALFIT model (right).  In addition to a Sersic exponential disk, three other compact components are required to obtain a fit without residuals:  two along the major axis displaced 7~kpc in projection from the center and an additional object close to the minor axis.}
\label{fig:deconv}
\end{figure}

The decomposition described above also functions as a basic deconvolution of the image near the galaxy.   In Figure \ref{fig:deconv} we show both the observed $i$-band image and its deconvolved equivalent, which is the sum of the model produced by GALFIT (\emph{not} re-convolved with the PSF) and the fit residuals.  The residuals are added both to accurately show the noise level, and also to avoid falsely removing any real signal that may be left in the residuals.

Deconvolution should always be approached with caution, and our modeling cannot identify faint, small-scale ($\ll 1$\arcsec) structures that are likely to be present or more complex structures such as bars or arcs.  Nevertheless, since the observed components are all separated on scales larger than the 0.8\arcsec\ seeing disk, we expect that they likely represent real substructure in the galaxy --- projected spiral arms or localized intense star-forming regions (the regular H$\alpha$ rotation curve in Figure \ref{fig:rotcurve} suggests they are not merging galaxies).  Further observations with better image quality (such as from the {\it Hubble Space Telescope}) would be necessary to unambiguously resolve the structure of this system.  Nevertheless, it is clear that the host is a morphologically complex but predominately disk-like galaxy. 

\subsection{Emission-Line Measurements}
\label{sec:emissionlines}

\begin{deluxetable*}{lllll}  
\tabletypesize{\small}
\tablecaption{Fluxes of Detected Emission Lines in G1 and G2}
\tablecolumns{5}
\tablehead{
\colhead{Galaxy} &
\colhead{Line Species} &
\colhead{Rest Wavelength} &
\colhead{Unsubtracted Flux\tablenotemark{a}} &
\colhead{Subtracted Flux\tablenotemark{b}}
\\
\colhead{} &
\colhead{} &
\colhead{(\AA)} & 
\colhead{($10^{-17}$ erg s$^{-1}$)} &
\colhead{($10^{-17}$ erg s$^{-1}$)}
}
\startdata
G1 & \OII\tablenotemark{c}& 3727     & $16.2 \pm 1.8$  &  $19.6 \pm 2.1$  \\
   & H$\beta$ & 4861.33              & $12.7 \pm 1.8$  &  $21.8 \pm 1.9$  \\
   & \OIII       & 5006.84           & $ 5.9 \pm 1.5$  &  $ 5.1 \pm 1.2$  \\
   & H$\alpha$  & 6562.82            & $108.6\pm 1.4$  &  $112.8\pm 1.6$  \\ 
   & \NII        & 6548.06           & $12.8 \pm 1.1$  &  $13.1 \pm 1.2$  \\
   & \NII        & 6583.57           & $46.4 \pm 1.2$  &  $45.7 \pm 1.3$  \\
   & \SII        & 6716.44           & $20.4 \pm 3.2$  &  $20.4 \pm 3.1$  \\
   & \SII        & 6730.82           & $14.1 \pm 1.8$  &  $13.6 \pm 1.7$  \\
\hline
G2 & \OII\tablenotemark{c}& 3727     & $6.0 \pm 0.8$  &  ---  \\
   & H$\beta$ & 4861.33              & $2.8 \pm 1.7$  &  ---  \\
   & \OIII       & 5006.84           & $3.1 \pm 0.4$  &  ---  \\
\enddata 
\tablenotetext{a}{Assuming a flat continuum, corrected for Galactic extinction.}
\tablenotetext{b}{After subtracting the model continuum, corrected for Galactic extinction.}
\tablenotetext{c}{Combined flux of both lines of the \OII\ doublet.}
\label{tab:lineflux}
\end{deluxetable*}

The clear detections of numerous, bright emission lines in the spectrum of G1 allow us to apply standard diagnostics of the SFR, extinction, and metallicity to this galaxy.

We first extract the emission-line fluxes by assuming a flat continuum with no underlying stellar absorption component.   Taking a small region around each emission line or line complex, we fit a Gaussian function (or, in the case of blended lines, several summed Gaussians) added to a linear component representing the underlying flux with the \texttt{mpfit} package within IDL.  For each line, the integrated flux and uncertainty are obtained from the fit. The results are presented in the third column of Table \ref{tab:lineflux}. 

While the strong H$\alpha$ line (combined with the modest \NII/H$\alpha$ and \OIII/H$\beta$ ratios which associate this line with star formation rather than AGN activity; \citealt{Baldwin+1981}) gives unambiguous evidence of rapid, recent star formation, several other indicators suggest that an older stellar component is also present. Specifically, while many of the higher-order Balmer lines fall in the wavelength region of the dichroic where sensitivity is relatively poor ($\sim 5550$--5790~\AA), we do see a dip at the location of H$\delta$ and lower-significance dips at the locations of H$\eta$ and H$\zeta$, as well as (possibly) Ca H\&K (inset of Figure \ref{fig:spectrumg1}).  Interestingly, we see neither emission nor absorption at the location of H$\gamma$ where sensitivity is good, suggesting that the emission and absorption are of comparable strength, resulting in a flat apparent spectrum (the instrument resolution of 7~\AA\ is comparable to the intrinsic width of the Balmer line).  We also see evidence for a strong Balmer break in the photometry (\S \ref{sec:sed}).

The presence of Balmer absorption features raises the prospect that the simple procedure above used to measure the emission-line fluxes may be systematically underestimating some line strengths.  The large equivalent width of H$\alpha$ suggests that it is unlikely to be affected by uncertainties in the underlying continuum by more than a few percent, but the weaker H$\beta$ line may be much more strongly affected, which would impact the spectroscopically derived extinction (and therefore the extinction-corrected SFR). Consequently, after deriving an estimate of the stellar continuum from population synthesis modeling (\S \ref{sec:sed}, below), we subtract the continuum flux given by that model and then repeat the line fits with the continuum removed.  The results are given in the fourth column of Table \ref{tab:lineflux}.

For galaxy G2, we only present the line fluxes using the basic linear continuum-subtraction model.  As the continuum trace of this galaxy is extremely weak, this is a good assumption at the level of uncertainty in our measurements.  The fluxes of the probable lines (note that H$\beta$ is blended with a strong night-sky line and not clearly present) are given in Table \ref{tab:lineflux}.

\subsection{Chemical Abundance Analysis}

The strong and unambiguous detection of numerous emission lines makes it possible to measure the metallicity of this short-GRB host galaxy; this has been successfully done for only a few other short-GRB hosts so far.  Metallicity is expected to influence not only the lives of stars but also the outcome of their deaths as different kinds of explosion; thus, the burgeoning field of metallicity studies for different kinds of explosive events from both observational (for a review see \citealt{Modjaz2011}) and theoretical (e.g., \citealt{Hirschi+2005,Yoon+2005}) perspectives has emerged over the last few years.

The nebular oxygen abundance 12 + log(O/H) is the canonical choice of metallicity indicator for studies of the interstellar medium.  Using the derived line fluxes from Table \ref{tab:lineflux} and correcting for host-galaxy reddening $E_{B-V}= 0.60 \pm 0.42$ mag assuming H$\alpha$/H$\beta = 2.86$ from Case B recombination \citep{Osterbrock1989}, we measure the nebular oxygen abundance of the host galaxy following the technique described by \cite{Modjaz+2008,Modjaz+2011}.   Using the scales from \citet{PP04} (PP04), we find 12 + log(O/H) = $8.74^{+0.02}_{-0.03}$ based on the \NII/H$\alpha$ diagnostic (PP04--N2), and 12 + log(O/H) = $8.81^{+0.10}_{-0.12}$ based on the \OIII/\NII (PP04--O3N2) prescription. Likewise, we find 12 + log(O/H) = $9.17^{+0.09}_{-0.11}$ on the scale of \citet{KD02} (KD02), and 12 + log(O/H) = $8.82^{+0.11}_{-0.16}$ on the scale of \citet{M91} (M91). Here we have computed the uncertainties in the measured metallicities by explicitly including the statistical uncertainties of the line-flux measurements, a conservative estimate of the uncertainty arising from continuum absorption contamination, and those in the derived host-galaxy reddening, and we propagate them into the metallicity determination. The ionization parameter $q$, iteratively derived in the KD02 models, is log$_{10}(q) = 7.59^{+0.4}_{-0.27}$.

Given the most recent estimate of the solar oxygen abundance [12 + log(O/H) = 8.69; \citealt{Asplund+2009}], together with the relative robustness of the $T_e$-based metallicity scale \citep{Bresolin+2009}, which the PP04--O3N2 scale is close to, we conclude the true value of the host metallicity $Z$ is most likely to be closer to the lower end of the estimated range, approximately 12 + log(O/H) = 8.8, or $Z = 1.1$\,${\rm Z}_{\odot}$; mildly super-solar.  However, most previous metallicity work on short GRB host galaxies \citepeg{Berger2009} has used the KD02 scale exclusively, which tends to produce systematically higher values than PP04 (see detailed discussions in \citealt{Kewley+2008,Moustakas+2010}  and references for possible reasons for the systematic offsets between different abundance diagnostics).  For more direct comparison with other GRB host galaxies, then, the KD02 value of 12 + log(O/H) = $9.17^{+0.09}_{-0.11}$ (a significantly larger value of  $Z = 3$\,${\rm Z}_{\odot}$) is more informative.  

In either case, we conclude that G1 is the most metal-rich host galaxy of any GRB (long or short) to date.  To our knowledge, the only super-solar long GRB host galaxies (as determined via emission spectroscopy, via the KD02 scale\footnote{Among the hosts discussed in this paragraph, only GRB 020819 was also measured on the PP04 scale: the reported value from \cite{Levesque+2010f} is $8.7 \pm 0.1$, similar to what we derive for GRB 100206A on the same scale.}) are the hosts of GRB\,020819 (12 + log(O/H) = $9.0 \pm 0.1$; \citealt{Levesque+2010f}), GRB\,050826 (12 + log(O/H) = $8.83 \pm 0.1$; \citealt{Levesque+2010g}) and GRB\,051022 (12 + log(O/H) $\approx$ 8.77; \citealt{Graham+2010,Graham+2011}).   Among short GRBs, the host galaxies of GRBs 051221A, 061210, and 070724A all have metallicities of 12 + log(O/H) = 8.8--8.9 (also using KD02; \citealt{Berger2009}) \cite{Kocevski+2010} report a KD02 value of 12 + log(O/H) = 9.1 for GRB 070724A.  Using the same scale, our measured value for GRB 100206A (12 + log(O/H) = $9.17^{+0.09}_{-0.11}$) identifies this galaxy as the most metal-rich GRB host known.

\begin{deluxetable*}{lllll}
\tabletypesize{\small}
\tablecaption{G1 Metallicity Determinations}
\tablecolumns{5}
\tablehead{
\colhead{Method} &
\colhead{Type\tablenotemark{a}} &
\colhead{Relevant lines\tablenotemark{b}} & 
\colhead{12 + log(O/H)} &
\colhead{Z/Z$_{\odot}$\tablenotemark{c}}
}
\startdata
PP04--N2       & Empirical & \NII, H$\alpha$                                    & $8.74^{+0.02}_{-0.03}$ & $1.12^{+0.05}_{-0.07}$ \\
PP04--O3N2     & Empirical  & \OIII, H$\beta$, \NII, H$\alpha$                   & $8.81^{+0.10}_{-0.12}$ & $1.32^{+0.34}_{-0.32}$ \\
M91 ($R_{23}$) & Theoretical  & \OII, \OIII, H$\beta$                              & $8.82^{+0.11}_{-0.16}$ & $1.35^{+0.87}_{-0.30}$ \\
KD02 (combined)& Theoretical  & \OII, \OIII, \NII, \SII, H$\beta$                  & $9.17^{+0.09}_{-0.11}$ & $3.02^{+0.87}_{-0.93}$  \\
Fitted\tablenotemark{d} & Theoretical  &\OII, \OIII, \NII, \SII, H$\alpha$, H$\beta$ & $9.04\pm0.09$          & $2.25\pm0.4$ \\
\enddata 
\tablenotetext{a}{~Principle of the metallicity calibration: theoretical (using photo-ionizaion models) or empirical (calibrated to observations of local star-forming regions.)}
\tablenotetext{b}{~Lines directly used as part of the metallicity determination prescription.  Most models additionally include H$\alpha$ and H$\beta$ to determine the host extinction, which is then used to correct the metal line fluxes.}
\tablenotetext{c}{~Gas-phase oxygen abundance (in Solar units, for log[O/H]$_\odot$ = 8.69).}
\tablenotetext{d}{~Weighted average of the lower three models in Table \ref{tab:models}; see \S \ref{sec:sed} for details.}
\label{tab:metallicity}
\end{deluxetable*}

\subsection{Spectral Energy Distribution}
\label{sec:sed}

\begin{deluxetable*}{llllll|lcc|l}  
\tabletypesize{\scriptsize}
\tablecaption{Results of Model Fits to Photometry and Spectroscopy of G1}
\tablecolumns{10}
\tablehead{
\colhead{Z/Z$_{\odot}$\tablenotemark{a}} &
\colhead{SFR$_0$\tablenotemark{b}} &
\colhead{SFR$_{\rm av}$\tablenotemark{c}} &
\colhead{Age\tablenotemark{d}} & 
\colhead{$\tau$\tablenotemark{e}} & 
\colhead{Mass\tablenotemark{f}} &
\colhead{Dust\tablenotemark{g}} & 
\colhead{$A_V$\,(old)\tablenotemark{h}} &
\colhead{$A_V$\,(young)\tablenotemark{i}} &
\colhead{$\chi^2$}
\\
\colhead{} &
\colhead{(M$_{\odot}$ yr$^{-1}$)} & 
\colhead{(M$_{\odot}$ yr$^{-1}$)} &
\colhead{(Gyr)} &
\colhead{(Myr)} &
\colhead{($10^9\,{\rm M}_{\odot}$)} & 
\colhead{law} & 
\colhead{(mag)} & 
\colhead{(mag)} & 
\colhead{}
}
\startdata
$2.56\pm0.51$ &\multicolumn{2}{c}{$11.6\pm 1.8$}& $ 11.5\pm 3.3$ & $\infty $ & $133\pm31 $ & smc   & \multicolumn{2}{c|}{$1.2\pm0.1$} & 25.47 / 14 \\
$2.16\pm0.46$ & $ 17.0\pm 3.0$ & $ 41.0\pm39.4$ & $  0.9\pm 0.8$ & $\infty $ & $36 \pm14 $ & smc   & \multicolumn{2}{c|}{$1.8\pm0.2$} & 17.00 / 13 \\
$2.58\pm0.36$ & $ 19.4\pm 4.2$ & $ 94.3\pm26.6$ & $  0.6\pm 0.2$ & $100    $ & $54 \pm5  $ & calz  & \multicolumn{2}{c|}{$1.8\pm0.2$} & 13.48 / 13 \\
$2.32\pm0.36$ &\multicolumn{2}{c}{$24.0\pm 6.3$}& $  2.4\pm 0.9$ & $\infty $ & $57 \pm16 $ & calz  & $ 1.3\pm0.1$ & $ 2.2\pm0.3$      & 12.82 / 13 \\
$1.94\pm0.75$ & $ 37.4\pm13.1$ & $ 18.9\pm14.0$ & $  3.4\pm 2.3$ & $\infty $ & $65 \pm21 $ & calz  & $ 1.1\pm0.2$ & $ 2.7\pm0.4$      &  9.16 / 12 \\
$2.27\pm0.46$ & $ 34.1\pm14.6$ & $ 82.0\pm24.4$ & $  0.6\pm 0.2$ & $100    $ & $52 \pm5  $ & calz  & $ 1.6\pm0.3$ & $ 2.5\pm0.5$      & 11.88 / 12 \\
\enddata 
\tablenotetext{a}{Gas-phase oxygen abundance (in Solar units, for log[O/H]$_\odot$ = 8.69).}
\tablenotetext{b}{Current SFR.}
\tablenotetext{c}{Average SFR.}
\tablenotetext{d}{Age of formation.}
\tablenotetext{e}{SFR decay time scale (for exponential-decline model).}
\tablenotetext{f}{Total stellar mass.}
\tablenotetext{g}{Best-fit dust extinction curve (Milky Way, Small Magellanic Cloud, or Calzetti).}
\tablenotetext{h}{Extinction for $t > 100$ Myr stars.}
\tablenotetext{i}{Extinction for $t < 100$ Myr stars.}
\label{tab:models}
\end{deluxetable*}

We have recently developed a simple, flexible code for fitting multiwavelength photometric observations of galaxies.  Implemented in IDL using \texttt{mpfit}, the code uses a small number of smoothly varying fundamental parameters (metallicity, mass, current SFR, and parameterizations of the past star-formation history) and a grid of population-synthesis models from \cite{bc03} to calculate the stellar spectral energy distribution (SED).  This is extinguished by dust using one of several standard extinction laws, and the absorbed energy is reradiated in the form of a multi-temperature dust graybody and mid-IR polycyclic aromatic hydrocarbon (PAH) emission features.  Emission lines are also included, scaled using the relations from \citet{Kennicutt1998}, \citet{KD02}, and \citet{Kewley+2004} using the current SFR, metallicity, and ionization.

We initially fit our models using only the broadband photometry, fixing the SFR and extinction using the H$\alpha$ and H$\beta$ emission-line flux measurements.  However, the probability of strong, unresolved stellar absorption underlying H$\beta$ makes it difficult to apply these constraints in a self-consistent way --- and, furthermore, the age/extinction degeneracy makes it nearly impossible to place useful constraints on the stellar population age without photometry in the vicinity of the Balmer and 4000~\AA\ breaks.  Fortunately, our flux-calibrated spectrum is of sufficient quality to fill these gaps.  Using this spectrum, we calculate synthetic photometry using a series of customized ``filters'' covering several independent, critical regions of the spectrum --- the major absorption and emission lines (\OII, \OIII\ $\lambda$5007, \NII\ $\lambda$6854, \SII\ $\lambda\lambda$6717, 6731, H$\delta$, H$\gamma$, H$\beta$, H$\alpha$) as well as interline continuum regions just blueward of the Balmer break (5650--5730~\AA\ and 5850--6050~\AA).  (With the exception of the \OII\ line, we do not attempt this technique blueward of the Balmer break due to the weak continuum trace and relatively large uncertainty in the overall flux calibration of the blue side.)   We exclude the W3 filter in these fits, since the model is subject to large systematic uncertainties in this wavelength region regarding the fraction of dust emission in PAH lines and the possible presence of highly embedded star formation.   (However, all of our models which reasonably fit the optical/NIR points also accurately predicted the W3 flux within 2$\sigma$, and including this filter does not qualitatively change any results.)

\begin{figure}
\centerline{
\includegraphics[scale=0.65,angle=0]{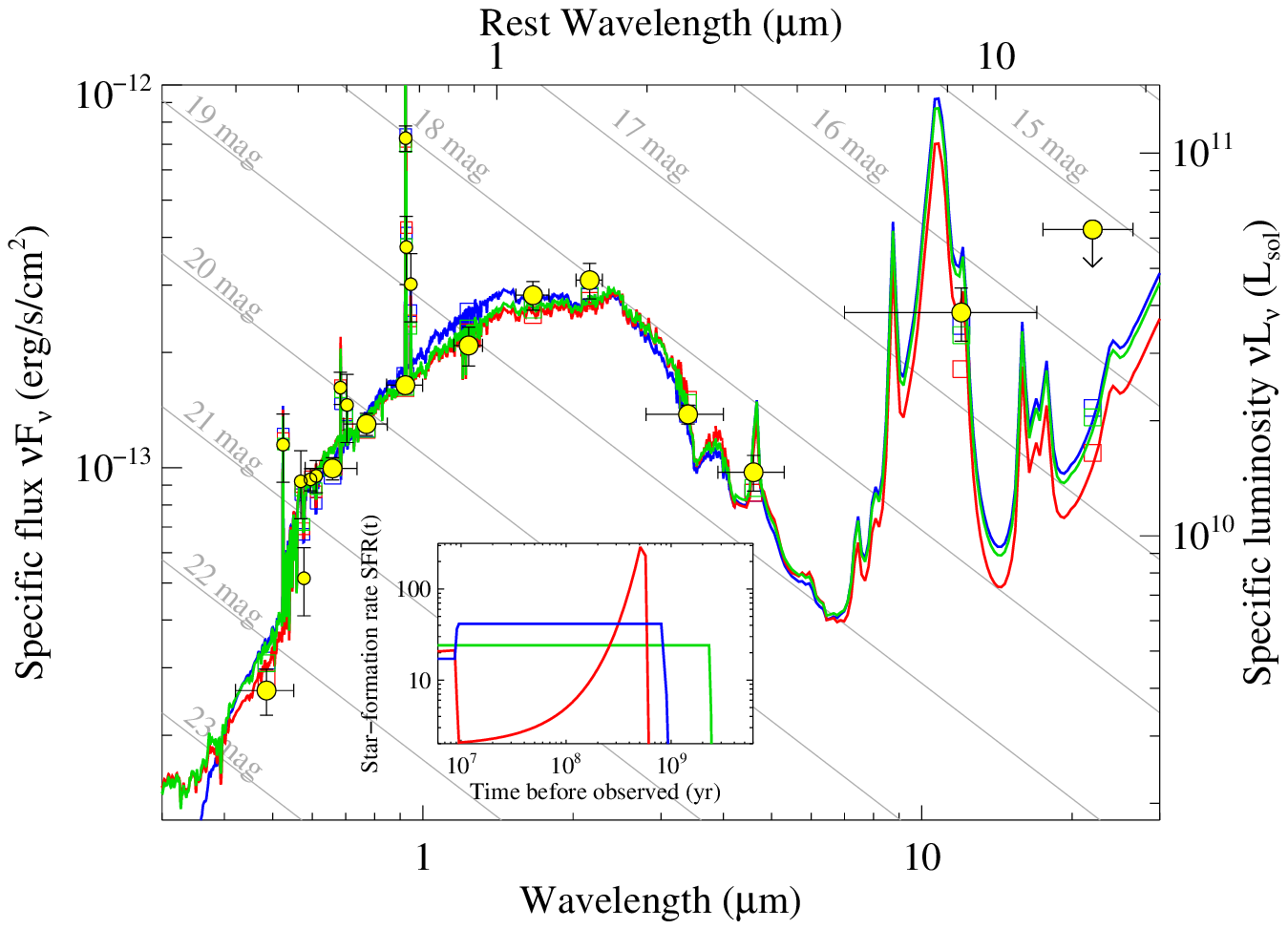}}  
\caption{Photometry of the putative host galaxy G1 of GRB 100206A, fit with stellar population models using our own implementation of the \cite{bc03} templates (including nebular lines, dust extinction, and mid/far-IR dust emission) for different assumptions of the star formation history.  The broadband photometry is supplemented by synthetic narrowband photometry of major emission and absorption-line regions (and interline regions near the Balmer break).  Three different star-formation history models are shown, all of which produce similar results (see also Table \ref{tab:models}).  The \emph{green} curve shows a strictly continuous star-formation history from the formation of the galaxy until the present time.  The \emph{blue} curve is also constant, except for an instantaneous change in the recent past. The \emph{red} curve assumes an impulsive star-formation episode at some point in the past with exponential decay time $\tau = 200$~Myr.  All three models use an age-dependent dust screen.  Broadband photometry is indicated with large yellow points; synthetic narrowband photometry is indicated with smaller points.  Empty colored squares show the synthetic fluxes for each filter.  The gray lines show contours of constant AB magnitude.}
\label{fig:fitsed}
\end{figure}

Three different models of the past star-formation history were attempted (see inset of Figure \ref{fig:fitsed}): purely constant (up to and including the present-day value), constant (with an instantaneous change at $t = 10^7$ yr), and exponentially falling with a 100~Myr e-folding time (also with a step at $t = 10^7$~yr). For each of these star-formation histories, we attempted two models of the dust attenuation: either a single dust screen applied to all stars uniformly, or with two different screens --- one applied to the young ($< 100$ Myr) stars and nebular lines, the other to older stars ($>100$ Myr).  We try Small Magellanic Cloud \citep{Gordon+2003}, Milky Way \citep{ccm}, and Calzetti \citep{Calzetti+2000} extinction ``laws'' for each fit (Local-Group type extinction as implemented using the generalized parameterization of \citealt{FitzpatrickMassa1990}) and report the result producing the best $\chi^2$ in Table \ref{tab:models}.

With the exception of the purely constant, single-extinction model (which produces a poor fit), all of these models produce reasonable fits to the combined broadband and pseudo-narrowband photometry of G1.   These different models are shown in Figure \ref{fig:fitsed}, with the parameters outlined in Table \ref{tab:models}.  Different assumptions naturally lead to an intrinsic dispersion of properties, but essentially all good fits share several features in common.   The current SFR is large but not extreme (20--40 M$_{\odot}$), and the stellar mass is also quite high, (3--7) $\times 10^{10}$ M$_{\odot}$.  The (maximum) age of the stellar population is at least 0.5 Gyr (at least 1 Gyr for the constant star-formation history).  The extinction is also large, $A_V = 1.8 \pm 0.2$ mag (for a single extinction) or $A_{V,{\rm young}} = 2.7 \pm 0.4$ mag and $A_{V,\rm{old}}=1.1\pm0.2$ mag (age-dependent extinction).  All models require a large bolometric luminosity of $L = (3-4) \times 10^{11} {\rm L}_{\odot}$, the large majority of which is emitted in the far-IR, thus classifying the galaxy as an unambiguous LIRG.  As we do not have any far-IR measurements directly sampling the thermal dust emission, it is possible that the luminosity is even higher than this if an additional deeply embedded ($A_V > 50$ mag) component is present in the galaxy.  However, the relatively modest mid-IR emission in the W3 filter and the W4 nondetection suggest that such an extra component is not likely to be present, as is the case for most high-redshift LIRGs \citep{Reddy+2010}, but unlike nearby LIRGs and ULIRGs \citepeg{Symeonidis+2008}.

G2 is only detected in four filters, and no hydrogen lines are evident in the spectrum, so it is not possible to constrain the properties of the galaxy in any detail.  Assuming no extinction, the galaxy is consistent with being a low-mass, non-starbursting galaxy with a modest current SFR of $\sim$0.5 M$_\odot$ yr$^{-1}$ and mass $\sim 4 \times 10^9 {\rm M}_\odot$. 

\section{Discussion}
\label{sec:disc}

\subsection{Association of G1 with GRB 100206A}
\label{sec:pchance}

\begin{deluxetable*}{llllll}
\tabletypesize{\small}
\tablecaption{Chance Alignment Parameters for G1 and G2}
\tablecolumns{6}
\tablehead{
\colhead{Galaxy} &
\colhead{Filter} &
\colhead{$m$\tablenotemark{a}} & 
\colhead{$n_{m < m_{\rm G1}}$} &
\colhead{$P_{\rm chance}$} &
\colhead{$P_{\rm chance, all SHBs}$}
\\
\colhead{} &
\colhead{} &
\colhead{(mag)} & 
\colhead{(deg$^{-1}$ mag$^{-1}$)} & 
\colhead{} & 
\colhead{} 
}
\startdata
G1 & $g$      & 22.33 & 5730 & 0.05   & 0.80 \\
   & $R$      & 20.38 & 3160 & 0.03   & 0.58 \\
   & $i$      & 20.09 & 2280 & 0.02   & 0.46 \\
   & $z$      & 19.67 & 2280 & 0.02   & 0.47 \\
   & $J$      & 18.16 & 2000 & 0.02   & 0.42 \\
   & $H$      & 17.03 & 820  & 0.007  & 0.20 \\
   & $K_s$    & 16.19 & 740  & 0.006  & 0.18 \\
   & 3.6\,$\mu$m& 15.74 & 860  & 0.008  & 0.21 \\
   & 4.5\,$\mu$m& 15.14 & 950  & 0.008  & 0.23 \\
   & W3       & 11.23 & 170  & 0.0015 & 0.04 \\
\hline
G2 & $g$      & 25.21 & $1.1 \times 10^5$ & 0.61 & 1.0 \\
   & $R$      & 24.19 & $7.5 \times 10^4$ & 0.48 & 1.0 \\
   & $i$      & 23.95 & $8.3 \times 10^4$ & 0.52 & 1.0 \\
   & $z$      & 23.62 & $1.0 \times 10^5$ & 0.58 & 1.0 \\
\enddata 
\label{tab:pchance}
\end{deluxetable*}

We have previously suggested that the probability of chance alignment of an IR-bright source with an XRT position is very low.  Here we quantify this calculation, and demonstrate that the case for associating GRB 100206A with the galaxy G1 is quite strong.

Figure 1 shows the current XRT error circle relative to our optical imaging.  Note that while the galaxy is \emph{centered} outside the XRT error circle, a significant amount of flux from the southern end of the disk is in fact contained within the XRT error circle.

Roughly following \cite{Bloom+2002}, an estimate of the probability of chance association $P_{\rm chance}$ can be provided by

$P_{\rm chance} = 1 - {\rm exp(}- A_{\rm assoc} \times n_{m < m_{\rm obj}} {\rm ).} $

\noindent Here, $A_{\rm assoc}$ is the area on the sky of the region in which an error circle centered in that region would still lead us to connect the GRB and the galaxy.  We take this as a circle of radius equal to the sum of the galaxy's major axis and the 90\% confidence radius of the XRT position: $\sim 6$\arcsec, so $A_{\rm assoc} = 113$ square arcseconds.

The term $n_{m < m_{\rm obj}}$ is the density on the sky of objects of equal or greater brightness.  Because of G1's unusual colors, this factor is strongly dependent on the filter chosen.  Rather than restricting our calculation to a single filter, then, we calculate $P_{\rm chance}$ for all filters in which the galaxy was detected.  The number density of galaxies on the sky for various filters was taken from a variety of sources:  \cite{Yasuda+2001} for the SDSS $giz$ bands; \cite{Hogg+1997} for the $R$ band; \cite{Jarrett2004,Frith+2006,Imai+2007,Maihara+2001} for the $J$, $H$, and $K$ bands.  For the W3 filter we use the number-count plots in \cite{Jarrett+2011}, while for the W1 and W2 filters we adopt the more precisely determined {\it Spitzer}/IRAC number counts in \cite{Fazio+2004}, transforming the WISE filters to equivalent IRAC fluxes using synthetic photometry of our model. Results were checked for consistency with the compilation graphs at http://astro.dur.ac.uk/$\sim$nm/pubhtml/counts/counts.html for relevant filters.

The sky densities and $P_{\rm chance}$ values are presented in Table \ref{tab:pchance}.  The fourth column lists the basic $P_{\rm chance}$ as described above, the value of which is small almost regardless of the filter examined --- ranging from a few percent in the optical filters down to $\sim 10^{-3}$ in the W3 filter, similar to the strength of association of GRB 050509B to its putative elliptical host (\citealt{Bloom+2006}, which was based on $r$-band photometry).

To adopt a more skeptical perspective, it should be noted that the number of localized short GRBs is no longer small, and curious chance alignments are bound to happen eventually.  We therefore also calculate the $P_{\rm chance}$ using the combined area covered by all 46 short GRBs with X-ray localizations to date --- $\sim 733$ square arcseconds, or (adding the 3\arcsec\ major-axis of the galaxy) 3600 square arcseconds.  The value of $P_{\rm chance}$ is 0.18--0.8 for the optical and NIR filters, an obviously less convincing result, indicating that the global short-GRB sample size is indeed approaching a point where bright foreground or background interloper galaxies can be expected somewhere in the full sample.  However, when comparing to the W3 filter magnitude --- where the unusual characteristics of this galaxy are most pronounced --- the probability is still quite low at only 0.04.   In other words, the current short-GRB sample size is still a factor of $\sim 5$--10 too small for chance alignment with a comparably mid-IR bright galaxy to be expected.

An alternative hypothesis that GRB 100206A originates from G2 was first presented by \cite{GCN10395}.  While this certainly cannot be ruled out, we judge it to be relatively unlikely given the faintness of this galaxy and the size of the XRT error circle: for all four filters, $P_{\rm chance}$ of the individual association is $\sim 0.5$ (i.e., a randomly positioned error circle of this size will enclose a comparable source approximately half the time).

Based on these considerations, we argue that the host galaxy of this event is G1.   As with all short-GRB host associations to date, this one is based entirely on {\it a posteriori} arguments and---given the absence of an absorption spectrum---is impossible to prove conclusively.  If this GRB is associated, however, the unique properties of this galaxy (among short-GRB hosts so far) have the potential to impact our understanding of the short-GRB progenitor in important ways.  In the remainder of the paper, we will assume that this association is correct and examine its implications.

\subsection{Characteristics of the Galaxy and Implications for the Progenitor}
\label{sec:implications}

With an SFR of at least 15 M$_\odot$ yr$^{-1}$ (and 30--40 M$_\odot$ yr$^{-1}$ in our best-fit models), G1 is the most rapidly star-forming host of a short-hard GRB to date.  The next-highest reported SFR among short-GRB hosts in the literature is only 6.1 M$_{\odot}$ yr$^{-1}$ (GRB 060801; \citealt{Berger2009}), much less than our inferred value.  (After that, the next-highest short-GRB SFRs belong to GRBs 061217 and 070724, both at a mere 2.5 M$_{\odot}$ yr$^{-1}$.)\footnote{While extreme, the \emph{qualitative} properties of G1 are not completely without precedent among short GRB hosts.  The host of GRB~070724 \citep{Berger+2009,Kocevski+2010} in particular, despite having an SFR and stellar mass an order of magnitude lower than those of G1, is also well above the average among short-GRB hosts in both these properties.  The host of GRB~070724 is also significantly dust-extinguished, and in fact has a dust-reddened optical afterglow.}  The intense ongoing star formation raises the possibility that the GRB itself may be the product of a short-lived star, coming from a young stellar progenitor produced at the same time as the large population of massive stars currently powering the bright nebular and mid-IR emission.  However, more detailed analysis of the galaxy suggests that a large population of evolved stars must be present as well, and that the current SFR is not even unusual relative to the rest of the galaxy's history. 

Unfortunately, there is no direct means of determining which of these two populations produced the GRB.  The X-ray position does not locate the burst within the host (and, in fact, leaves open the possibility of a halo origin), nor can any properties of the absorbing column be inferred from the afterglow.  

Bereft of any direct means of evaluating the age of the progenitor---and, thus far, only a single example of a dusty, rapidly star forming, short-GRB host to date---we must resort again to statistical arguments.   Relative to the Universe as a whole, is this galaxy more likely to produce explosions from old progenitors, or young ones?   To phrase this question more precisely (under the simplifying assumption that the short-GRB rate is directly proportional to the number of stars of the appropriate age range and ignoring secondary effects such as detection biases), we can ask the following:  \emph{as a relative fraction} of all stars in the Universe at that redshift, are there more ``old'' stars or ``young'' stars in LIRG-like hosts at $z \approx 0.4$\ ?   (For now, we will encapsulate the ``unusual'' characteristics of this host galaxy entirely in its designation as a LIRG.)  Fortunately, this type of question has been addressed in great detail by recent studies:  the sites of star formation, the build-up of stellar mass, and the role of the LIRG phase in galaxy evolution have been major foci of recent work in extragalactic astronomy.  

We will first address the fraction of \emph{star formation} occurring in LIRGs.  At higher redshifts of $z>1$, a quite significant fraction of all star formation occurs in luminous, dusty galaxies (as much as 70\%; \citealt{LeFloch+2005}).  However, this fraction begins to fall rapidly below $z=1$ and it is already relatively low by $z=0.4$: Figure 14 of \citet{LeFloch+2005} indicates that about 20\% of star formation at that era occurs in LIRGs, and a consistent value of 10--20\% is given by \citet{PerezGonzalez+2005}.   It is therefore not unexpected that \emph{if} a small fraction of short GRBs were produced by young ($<$100 Myr) stars, we might have observed one in a LIRG --- though it is likely that several other short GRBs would be observed to occur in more ``typical'' star-forming galaxies first.   (Whether this has in fact occurred is a matter of debate: see \S \ref{sec:intro}.)

The fraction of all \emph{stellar mass} in LIRGs at moderate redshift is given by \cite{Caputi+2006}, who estimate that 24\% of the stellar mass at $z = 0.5$--1.0 is in LIRGs and ULIRGs.  The fraction at lower redshifts is not given, but Figure 8 of \citet{Salim+2009} suggests it is similar (within a factor of a few) to that at $z \approx 0.5$--1.0; the number of LIRGs does not begin to fall precipitously to the local density until about $z \lesssim 0.3$, when they essentially disappear in deep surveys.  The fraction of all stars cosmologically that are present in LIRGs in this redshift range is therefore about 10--20\%, quite comparable to the cosmological fraction of young stars in LIRGs.  Even by {\it a posteriori} arguments, then, we are unable to associate this burst with a specific stellar population age with any confidence.  

We can also examine the galaxy's \emph{morphological} properties for clues about the GRB's origin.  At $z \approx 0$, for example, LIRGs are predominantly or entirely the products of major mergers: transient and (relatively) rare phenomena in the life of a galaxy with clear observational signatures, such as nuclear starbursts and tidal features.   (This is not the case at $z \approx 1$, when merging systems become a minority among LIRGs/ULIRGs; \citealt{Zheng+2004}.) Independent of the SED, then, a GRB from within a merger system could be seen as an indication of association with a progenitor with age similar to the merger itself, on the grounds that the explosion of an older star by chance during the short-lived merger phase would be an unlikely coincidence.  No such signatures are present; while our resolution is not sufficient to robustly rule out a merger remnant, the majority of the galaxy's flux appears to be associated with an ordinary exponential disk.   The galaxy \emph{is} slightly asymmetric, in that the region of maximum flux is displaced by $\sim 1$ kpc from the center of the outer isophotes, but this does not appear to reflect the detailed properties of the galaxy.  Similarly, there are no obvious disturbances in the structure of the rotation curve (Figure \ref{fig:rotcurve}).  

Finally, we can examine the \emph{physical} characteristics of the galaxy from our SED modeling.  Evidence for recent, significant elevation in the SFR above its historical average (equivalently, a large specific star-formation rate SFR/M$_{*} \gg 1/t$) would also be indicative that the GRB caught this galaxy in an unusual, transient part of its history, which is much more likely for a young progenitor than an older one.  But again, there is no such evidence:  the current and past SFRs are comparable, and the specific SFR is a relatively unremarkable (SFR/M$_{*}$ $\approx$0.5 Gyr$^{-1}$).  This is the largest value for a short GRB host yet (SFR/M$_{*}$ ranges between 0--0.2 Gyr$^{-1}$ using the SFRs from \citealt{Berger2009} and masses from \citealt{Leibler+2010}), but only by a factor of 2--3.  The value does overlap the distribution of (massive-star associated) long GRB hosts (typical values of SFR/M$_{*}$ range from 0.1--50 Gyr${-1}$, \citealt{Savaglio+2009,CastroCeron+2010}), but only at the low end: it is not an extreme value indicating an unambiguously star-bursting galaxy.

All available evidence, therefore, suggests that this is a relatively ordinary $z \approx 0.4$ galaxy when weighted either by star formation or by stellar mass.  Furthermore, it is (probably) being observed at a fairly typical moment in its history, with no clear indication of a large, recent increase in the SFR related to a merger or other transient event.  The properties of the galaxy give no clear means of distinguishing a short-lived from a long-lived progenitor of this system.

\section{Conclusions}
\label{sec:conc}

Seen as an isolated event, the age of the progenitor of GRB 100206A is frustratingly inconclusive:  essentially every diagnostic we have applied to the host system is consistent with both a short-lived progenitor ($<$10 Myr) and a long-lived one ($>1$ Gyr).  While the galaxy has a large current SFR ($\sim 30$ M$_\odot$ yr$^{-1}$), this is matched by a substantial pre-existing stellar mass ($>10^{11}$ M$_\odot$).  This, and a largely ordinary disk-like morphology, suggest a steady star-formation history and give no direct indication of the progenitor's age.

Seen in the broader context of other short GRB host-galaxy work, however, our conclusions can be seen as generally supportive of the prevailing paradigm: as discussed in \S \ref{sec:implications}, the significant amount of stellar mass present in LIRG systems at $z>0.3$ suggests that it should be no surprise that a few percent of short GRBs should occur within these massive and actively star-forming galaxies, even if their progenitor is exclusively a long-lived object such as a neutron-star binary.   The fraction should be even higher if some of the detection biases are considered: preferential ejection of the progenitor system from low-mass host galaxies (due to supernova ``kicks''), as well as preferential detection of X-ray or optical afterglow from systems in dense environments, would both have much less impact on massive galaxies such as G1 than on more typical lower-mass galaxies (which would be actively selected against), favoring the discovery of this type of host.  Given what has become a quite large short-GRB  sample (46 objects with X-ray or optical localizations and at least 20 likely host galaxies), the occurrence of an event in a LIRG is not only unsurprising for an older progenitor, but could have been predicted.

This, coupled to the lack of \emph{any} definitive signature tying GRB 100206A to recent star formation in its host galaxy despite careful scrutiny of several possible lines of argument (including the SED-inferred star-formation history and morphology), can be interpreted as evidence that this event is the product of a relatively mature progenitor system in the galaxy.   Given the precedent set by other short GRBs, the most natural explanation for this burst is that it was produced in the same manner as the rest of the population --- for which a compelling statistical case is now building for a  generally old ($>100$ Myr) progenitor, even if the delay-time distribution is indeed broad.

Of course, a younger progenitor is not ruled out.  If this event \emph{were} the result of a young progenitor (and especially a massive star), it is interesting to note that the high metallicity of this system---at least Solar and perhaps significantly higher---would nevertheless be indicative of an origin distinct from that of long GRBs, which quite notably avoid dusty, massive LIRG-like systems in the local universe.  In that case this event would be more indicative of a broad delay-time distribution for the short-GRB population, rather than an origin from the same population that produces LGRBs due to, for example, off-axis collapsars \citep{Lazzati+2010}.   While it is possible that a few long-duration GRBs may occur in high-metallicity regions \citep{Levesque+2010f}, these appear to be relatively rare \citep{Modjaz+2008,Stanek+2006}, so \emph{even if} this event were to be associated with the youngest stellar population in this galaxy, it would nevertheless continue to support a distinction between the progenitor systems of short and long GRBs.

\vskip 1cm

\acknowledgments

Support for this work was provided by NASA through Hubble Fellowship
grant HST-HF-51296.01-A awarded by the Space Telescope Science
Institute, which is operated by the Association of Universities for
Research in Astronomy, Inc., for NASA, under contract NAS 5-26555.
A.V.F. and S.B.C. acknowledge generous support from Gary and Cynthia
Bengier, the Richard and Rhoda Goldman Fund, NASA/\Swift\ grants
NNX10AI21G and GO-7100028, and NSF grant AST--0908886.

PAIRITEL is operated by the Smithsonian Astrophysical Observatory
(SAO) and was made possible by a grant from the Harvard University
Milton Fund, a camera loan from the University of Virginia, and
continued support of the SAO and UC Berkeley. The PAIRITEL project is
further supported by NASA/{\it Swift} Guest Investigator grant
NNX08AN84G and NNX10AI28G. We acknowledge support from the NSF/AAG 1009991.

We wish to acknowledge N~.R. Tanvir, A.~J. Levan, E. Berger, and R. Chornock,
as well as the Gemini-North staff (in particular T.~Geballe, M.~Lemoine-Busserolle, 
S.~Cote, and R.~Mason) for acquisition of the Gemini data.  We also heartily thank J.~M.
Silverman for acquisition of the second epoch of Keck spectroscopy.
The W. M. Keck Observatory is operated as a scientific partnership among the
California Institute of Technology, the University of California, and
NASA.  The Observatory was made possible by the generous financial
support of the W. M. Keck Foundation.  We wish to extend special
thanks to those of Hawaiian ancestry on whose sacred mountain we are
privileged to be guests.

This work made use of data supplied by the UK \Swift\ Science Data
Centre at the University of Leicester.  It is a pleasure to thank all
members of the \Swift\ team, who built and continue to operate this
successful mission. This publication makes use of data products from 
the Wide-field Infrared Survey Explorer, which is a joint project of 
the University of California, Los Angeles, and the Jet Propulsion 
Laboratory/California Institute of Technology, funded by the National 
Aeronautics and Space Administration.  This research also made use of the NASA/IPAC
Extragalactic Database (NED), which is operated by the Jet Propulsion
Laboratory, California Institute of Technology, under contract with
NASA.  The National Energy Research Scientific Computing Center, which 
is supported by the Office of Science of the U.S. Department of Energy 
under Contract No. DE-AC02-05CH11231, provided staff, computational, 
research, and data storage in support of this project.

D.~P. acknowledges useful discussions with Evan Kirby, Vivian U,
and Enrico Ramirez-Ruiz, and helpful comments on the manuscript from 
D.~A. Kann.

\bigskip

{\it Facilities:} \facility{Swift}, \facility{Keck:I (LRIS)},
\facility{PAIRITEL}, \facility{Gemini:North (GMOS)}, \facility{WISE}


\end{document}